\documentstyle[12pt]{article}
\newcommand{\GL}{ G_{\rm L}}
\newcommand{\GR}{ G_{\rm R}}
\newcommand{\FL}{ F_{\rm L}}
\newcommand{\FR}{ F_{\rm R}}

\begin{document}

\title{THE SPHALERON BARRIER IN THE PRESENCE OF FERMIONS}
\vspace{1.5truecm}
\author{
{\bf Guido Nolte}\\
Fachbereich Physik, Universit\"at Oldenburg, Postfach 2503\\
D-26111 Oldenburg, Germany
\and
{\bf Jutta Kunz}\\
Instituut voor Theoretische Fysica, Rijksuniversiteit te Utrecht\\
NL-3508 TA Utrecht, The Netherlands\\
and\\
Fachbereich Physik, Universit\"at Oldenburg, Postfach 2503\\
D-26111 Oldenburg, Germany}

\vspace{1.5truecm}

\date{August 18, 1993}

\maketitle
\vspace{1.0truecm}

\begin{abstract}
We calculate the minimal energy path over the sphaleron
barrier in the pre\-sen\-ce of fermions,
assuming that the fermions of a doublet are degenerate in mass.
This allows for spherically symmetric ans\"atze
for the fields, when the mixing angle dependence
is neglected.
While light fermions have little influence on the barrier,
the presence of heavy fermions ($M_F \sim$ TeV) strongly
deforms the barrier, giving rise to additional
sphalerons for very heavy fermions
($M_F \sim$ 10 TeV).
Heavy fermions form non-topological solitons
in the vacuum sector.
\end{abstract}
\vfill
\noindent {Utrecht-Preprint THU-93/17} \hfill\break
\vfill\eject

\section{Introduction}

In 1976 't Hooft [1] observed that
the standard model does not absolutely conserve
baryon and lepton number
due to the Adler-Bell-Jackiw anomaly.
The process 't Hooft considered was
spontaneous fermion number violation due to instanton
induced transitions.
Fermion number violating
tunnelling transitions between
topologically distinct vacua might indeed
be observable at high energies at future accelerators [2,3].

Manton considered
the possibility of fermion number
violation in the standard model
from another point of view [4].
Investigating the topological structure
of the configuration space of the Weinberg-Salam theory,
Manton showed that there are noncontractible
loops in configuration space, and
predicted the existence of a static, unstable solution
of the field equations,
a sphaleron [5], representing
the top of the energy barrier between
topologically distinct vacua.

At finite temperature this energy barrier between
topologically distinct vacua can be overcome
due to thermal fluctuations of the fields,
and fermion number violating
vacuum to vacuum transitions
involving changes of baryon and lepton number
can occur.
The rate for such baryon number violating processes
is largely determined by a Boltzmann factor,
containing the height of the barrier at a given
temperature and thus the energy of the sphaleron.
Baryon number violation in the standard model due to
such transitions over the barrier may be relevant
for the generation of the baryon asymmetry of the universe
[6-10].

While the barrier between topologically distinct vacua
is traversed,
the Chern-Simons number changes continuously
from $N_{\rm CS}$ in one vacuum sector
to $N_{\rm CS}+1$ in the neighbouring vacuum sector.
At the same time one occupied fermion level
crosses from the positive continuum
to the negative continuum,
leading to the change in fermion number.
When considered in the background field approximation
this level crossing phenomenon
predicts the existence of a fermion zero mode
precisely at the top of the barrier, at the sphaleron.
For massless fermions this zero mode
is known analytically [11-13].

Considering the minimal energy path over the barrier [14,15],
the fermionic level crossing was demonstrated recently
in the background field approximation
for the Weinberg-Salam theory [16],
assuming that the fermions of a doublet are degenerate in mass.
This assumption, violated in the standard model,
allows for spherically symmetric ans\"atze
for all of the fields, when the mixing angle dependence
is neglected (which is an excellent approximation [17,18]).

With these assumptions we here calculate the
minimal energy path over the barrier self-consistently,
varying both the fermion mass and the Higgs mass.
The presence of the fermions influences
the energy and the shape of the barrier,
which need no longer be symmetric with respect to
the sphaleron. Neither must the fermion zero mode
occur precisely at the top of the barrier.

While our main concern is the study of the barrier
for light fermions up to about the top quark mass,
we also consider heavy fermions with masses in the TeV
region, because the possibility of rapid anomalous decay
was discussed recently for heavy fermions,
and the existence of a new type of soliton was conjectured [19],
which is distinct from the known weak non-topological
soliton [19-21] representing a weak chiral soliton.
Even for heavy fermions we employ the valence fermion
approximation, neglecting all boson loops and the
effects of the Dirac sea (as in [19]).

We briefly review in section 2
the Weinberg-Salam lagrangian
and the anomalous currents
for vanishing mixing angle
and for degenerate fermion doublets.
In section 3 we present our ansatz,
obtain the energy functional and the Chern-Simons number
and derive the equations of motion.
In section 4 we present our results.
First we discuss the influence of the presence of fermions
on the sphaleron. Then we exhibit the change
in energy and shape of the barrier
and present the fermion eigenvalue along the
minimal energy path.
We end this section with a discussion of the
non-topological soliton,
which, for heavy fermions,
is reached at the end of the path instead of the vacuum.
We present our conclusions in section 5.

\section{\bf Weinberg-Salam lagrangian}

We consider the bosonic sector of the Weinberg-Salam theory
in the limit of vanishing mixing angle.
In this limit the U(1) field
decouples and can consistently be set to zero
\begin{equation}
{\cal L}_{\rm b} = -\frac{1}{4} F_{\mu\nu}^a F^{\mu\nu,a}
+ (D_\mu \Phi)^{\dagger} (D^\mu \Phi)
- \lambda (\Phi^{\dagger} \Phi - \frac{1}{2}v^2 )^2
\   \end{equation}
with the SU(2)$_{\rm L}$ field strength tensor
\begin{equation}
F_{\mu\nu}^a=\partial_\mu V_\nu^a-\partial_\nu V_\mu^a
            + g \epsilon^{abc} V_\mu^b V_\nu^c
\ , \end{equation}
and the covariant derivative for the Higgs field
\begin{equation}
D_{\mu} \Phi = \Bigl(\partial_{\mu}
             -\frac{1}{2}ig \tau^a V_{\mu}^a \Bigr)\Phi
\ . \end{equation}
The ${\rm SU(2)_L}$
gauge symmetry is spontaneously broken
due to the non-vanishing vacuum expectation
value $v$ of the Higgs field
\begin{equation}
    \langle \Phi \rangle = \frac{v}{\sqrt2}
    \left( \begin{array}{c} 0\\1  \end{array} \right)
\ , \end{equation}
leading to the boson masses
\begin{equation}
    M_W = M_Z =\frac{1}{2} g v \ , \ \ \ \ \ \
    M_H = v \sqrt{2 \lambda}
\ . \end{equation}
We employ the values $M_W=80 \ {\rm GeV}$, $g=0.65$.

For vanishing mixing angle,
considering only fermion doublets degenerate in mass,
the fermion lagrangian reads
\begin{eqnarray}
{\cal L}_{\rm f} & = &
   \bar q_{\rm L} i \gamma^\mu D_\mu q_{\rm L}
 + \bar q_{\rm R} i \gamma^\mu \partial_\mu q_{\rm R}
   \nonumber \\
           & - & f^{(q)} \bar q_{\rm L}
           (\tilde \Phi u_{\rm R} + \Phi d_{\rm R})
               - f^{(q)} (\bar d_{\rm R} \Phi^\dagger
                     +\bar u_{\rm R} \tilde \Phi^\dagger)
           q_{\rm L}
\ , \end{eqnarray}
where $q_{\rm L}$ denotes the lefthanded doublet
$(u_{\rm L},d_{\rm L})$,
while $q_{\rm R}$ abbreviates the righthanded singlets
$(u_{\rm R},d_{\rm R})$,
with covariant derivative
\begin{equation}
D_\mu q_{\rm L} = \Bigl(\partial_{\mu}
             -\frac{1}{2}ig \tau^a V_{\mu}^a \Bigr) q_{\rm L}
\ , \end{equation}
and with $\tilde \Phi = i \tau_2 \Phi^*$.
The fermion mass is given by
\begin{equation}
M_F=\frac{1}{\sqrt{2}}f^{(q)} v
\ . \end{equation}

Due to the U(1) anomaly
baryon number and lepton number are not conserved
\begin{equation}
\partial^\mu j_\mu^{\rm B,L}=-f_{\rm g} \partial^\mu K_\mu
\ ,  \end{equation}
where
\begin{equation}
 K_\mu=\frac{g^2}{16\pi^2}\varepsilon_{\mu\nu\rho\sigma} {\rm Tr}(
 {\cal F}^{\nu\rho}
 {\cal V}^\sigma
 + \frac{2}{3} i g {\cal V}^\nu {\cal V}^\rho {\cal V}^\sigma )
\   \end{equation}
(${\cal F}_{\nu\rho} = 1/2 \tau^i F^i_{\nu\rho}$,
${\cal V}_\sigma = 1/2 \tau^i V^i_\sigma$)
is the Chern-Simons current and $f_{\rm g}$ is the number of
generations.
In the unitary gauge the topological baryon number $Q_{\rm B}$,
carried by a configuration, is determined
by its Chern-Simons number $N_{\rm CS}$,
\begin{equation}
N_{\rm CS} = \int d^3r K^0
\ . \end{equation}
For the vacua the Chern-Simons number is identical to the
integer winding number,
while the sphaleron at the top of the barrier carries
half integer Chern-Simons number [5].

\section{\bf Equations of Motion}

The sphaleron barrier can be obtained by
constructing a family of field configurations
for the fermionic, gauge and Higgs fields,
which interpolates smoothly from one vacuum sector to another,
minimizing the energy along this path
as a function of the Chern-Simons number [14].
Employing a spherically symmetric ansatz for the fields,
we now derive the equations of motion, the energy functional
and the Chern-Simons number.

\subsection{\bf Ansatz}

In the limit of vanishing mixing angle
the general static, spherically symmetric ansatz for the
gauge and Higgs fields is given by [22]
\begin{eqnarray}
  V_i^a & = & \frac{1-f_A(r)}{gr} \varepsilon_{aij}\hat r_j
  + \frac{f_B(r)}{gr} (\delta_{ia}-\hat r_i \hat r_a)
  + \frac{f_C(r)}{gr} \hat r_i \hat r_a  \ , \\
  V_0^a & = & 0\ , \\
    \Phi & = & \frac{v}{\sqrt {2}}
  \Bigl(H(r) + i \vec \tau \cdot \hat r K(r)\Bigr)
    \left( \begin{array}{c} 0\\1  \end{array} \right)
\ , \end{eqnarray}
and involves the five radial functions
$f_A(r)$, $f_B(r)$, $f_C(r)$,
$H(r)$ and $K(r)$.

To retain spherical symmetry
we consider only fermion doublets degenerate in mass.
The corresponding spherically symmetric ansatz for
the fermion eigenstates is the hedgehog ansatz,
\begin{equation}
q_{\rm L}(\vec r\,,t) = e^{-i\omega t} M_W^{\frac{3}{2}}
\bigl[ G_{\rm L}(r)
+ i \vec \sigma \cdot \hat r F_{\rm L}(r) \bigr] \chi_{\rm h}
\ , \end{equation}
\begin{equation}
q_{\rm R}(\vec r\,,t) = e^{-i\omega t} M_W^{\frac{3}{2}}
\bigl[ G_{\rm R}(r)
- i \vec \sigma \cdot \hat r F_{\rm R}(r) \bigr] \chi_{\rm h}
\ , \end{equation}
where the normalized hedgehog spinor $\chi_{\rm h}$
satisfies the spin-isospin relation
\begin{equation}
\vec \sigma \chi_{\rm h} + \vec \tau \chi_{\rm h} = 0
\ . \end{equation}

The ansatz is form-invariant under
a spherically symmetric gauge transformation with
the unitary matrix
\begin{equation}
U(\vec r\,)=\exp(i\frac{\Theta(r)}{2} \vec\tau \cdot \hat r)
\ . \end{equation}
Under such a U(1) transformation the functions transform
as follows
\begin{eqnarray}
f_A+if_B &\longrightarrow& \exp(i\Theta)(f_A+if_B) \ , \nonumber\\
f_C &\longrightarrow& f_C+r\Theta'  \ , \nonumber\\
H+iK &\longrightarrow&  \exp(i\frac{\Theta}{2})(H+iK) \ , \nonumber\\
\FL+i\GL &\longrightarrow&  \exp(i\frac{\Theta}{2})(\FL+i\GL) \ , \nonumber\\
\FR+i\GR &\longrightarrow&  \FR+i\GR \ .
\end{eqnarray}

\subsection{\bf Energy functional}

The ansatz eqs.(12)-(17) leads to the
spherically symmetric energy functional
\begin{equation}
E = E_{\rm b} + E_{\rm f}
\ , \end{equation}
with the bosonic part
\begin{eqnarray}
 E_{\rm b} & = & \frac{4\pi M_W}{g^2} \int^{\infty}_0 dx
         \Bigl[  \frac{1}{2x^2} (f^2_A + f^2_B  -1)^2
         + (f'_A + \frac{f_Bf_C}{x})^2
         + (f'_B - \frac{f_Af_C}{x})^2
\nonumber \\
   & + & (K^2+H^2) (1+f_A^2+f^2_B +
         \frac{f_C^2}{2})+2f_A (K^2-H^2) - 4f_B H K
\nonumber \\
   & - & 2xf_C (K'H - KH') + 2x^2(H'^2+K'^2)
      + \epsilon x^2 (H^2+K^2 -1)^2
    \Bigr]
\ , \end{eqnarray}
where
\begin{equation}
\epsilon = \frac{4\lambda}{g^2}
         = \frac{1}{2}\Bigl( \frac{M_H}{M_W} \Bigr)^2 \nonumber
\ , \end{equation}
and the fermionic part
\begin{eqnarray}
 E_{\rm f} & = & 4\pi M_W \int^{\infty}_0 dx x^2
         \Bigl[  \FR'\GR-\GR'\FR+\frac{2}{x}\FR\GR
         +    \FL'\GL-\GL'\FL+\frac{2}{x}\FL\GL
\nonumber \\
&-& 2 \frac{1-f_A}{x}\GL\FL
+\frac{f_B}{x} (\GL^2-\FL^2)
+\frac{f_C}{2x} (\GL^2+\FL^2)
\nonumber \\
&+&      2\tilde{M}_F H(\GR\GL-\FR\FL)
       - 2\tilde{M}_F K(\FR\GL+\FL\GR)
\Bigr]
\ . \end{eqnarray}
We have introduced the dimensionless coordinate
\begin{equation}
x=M_Wr \nonumber
\   \end{equation}
and expressed the fermion mass in units of $M_W$
\begin{equation}
\tilde{M}_F= M_F/M_W \nonumber
\ . \end{equation}
The energy functional is invariant under the U(1) gauge
transformation {eq.(18)}.

The fermion functions entering the energy functional
need to be normalized.
When $N$ fermions occupy the eigenstate eqs.(15)-(16)
the normalization condition is
\begin{equation}
         4\pi  \int^{\infty}_0 dx x^2
(\GR^2+\FR^2+\GL^2+\FL^2)=N
\ . \end{equation}

To construct the minimal energy path
over the sphaleron barrier, we then consider the functional
\begin{equation}
W=E -\omega N + \frac{8 \pi^2 M_W} {g^2} \xi N_{\rm CS}
\ , \end{equation}
where $\omega$ is the fermion eigenvalue and
$\xi$ is a dimensionless lagrange multiplier [14].
The Chern-Simons number of a given configuration is
\begin{equation}
N_{\rm CS} = \frac{1}{2\pi} \int^{\infty}_0 \ dx \Bigl[
(f_A^2+f^2_B) (\frac{f_C}{x} - \theta') -
(\frac{f_C}{x} - \Theta') -
\Bigl(\sqrt {(f_A^2+f^2_B)}\
\sin (\theta - \Theta)\Bigr)'\ \Bigr]
\ , \end{equation}
where we have defined the function
$\theta(x) = {\rm arctan}({f_B}/{f_A})$.
The function $\Theta(x)$ is an arbitrary radial function,
associated with a general U(1) gauge transformation (18).
 From the expression (28)
the Chern-Simons number is readily obtained in an arbitrary gauge,
and in particlular in the gauge, where the identification
with the topological baryon number can be made.

\subsection{\bf Gauge choice}

We perform the calculations in the radial gauge, where $f_C=0$.
In this gauge the spatial part of the Chern-Simons current
contributes to the topological baryon number.
We therefore rotate to the unitary gauge,
where only the Chern-Simons number
determines the topological baryon number.
The corresponding gauge transformation
involves the function $\Theta(x)$,
which satisfies $\Theta(0)=0$ and
$\Theta(\infty)=\theta(\infty)$.
We then find for the Chern-Simons number the expression
\begin{equation}
N_{\rm CS}=
\frac{1}{2\pi}\int_0^{\infty}dx
(f_B f_A'-f_A f_B') +\frac{\theta(\infty)}{2\pi}
\ . \end{equation}

\subsection{\bf Equations}

Variation of the functional eq.(27)
leads to the following set of equations for the bosons
\begin{equation}
f_A''=\frac{f_A}{x^2}(f_A^2+f_B^2-1)+f_A(K^2+H^2)
+K^2-H^2+g^2x\FL\GL-\xi f_B'
\ , \end{equation}
\begin{equation}
f_B''=\frac{f_B}{x^2}(f_A^2+f_B^2-1)+f_B(K^2+H^2)
-2HK+\frac{1}{2}g^2x(\GL^2-F_L^2)+\xi f_A'
\ , \end{equation}
\begin{eqnarray}
H''&=& -\frac{2}{x}H'
+\frac{H}{2x^2}\Bigl((1-f_A)^2+f_B^2\Bigr) -\frac{K}{x^2}f_B
+\epsilon (H^2+K^2-1)H
\nonumber \\
&+& \frac{g^2 \tilde M_F}{2}(\GR\GL-\FR\FL)
\ , \end{eqnarray}
\begin{eqnarray}
K''&=& -\frac{2}{x}K'
+\frac{K}{2x^2}\Bigl((1+f_A)^2+f_B^2)\Bigr) -\frac{H}{x^2}f_B
+\epsilon (H^2+K^2-1)K
\nonumber \\
&-& \frac{g^2 \tilde M_F}{2}(\FR\GL+\FL\GR)
\ , \end{eqnarray}
and for the fermions
\begin{equation}
\tilde \omega G_{\rm L} - F'_{\rm L} - \frac{2}{x}F_{\rm L}
+\frac{1-f_A}{x} F_{\rm L}
-\frac{f_B}{x} G_{\rm L}
+\tilde M_F(-H G_{\rm R} + K F_{\rm R}) = 0
\ , \end{equation}
\begin{equation}
\tilde \omega F_{\rm L} + G'_{\rm L}
+\frac{1-f_A}{x} G_{\rm L}
+\frac{f_B}{x} F_{\rm L}
+\tilde M_F(H F_{\rm R} + K G_{\rm R}) = 0
\ , \end{equation}
\begin{equation}
-\tilde \omega G_{\rm R} + F'_{\rm R} + \frac{2}{x}F_{\rm R}
+\tilde M_F(H G_{\rm L} - K F_{\rm L}) = 0
\ , \end{equation}
\begin{equation}
\tilde \omega F_{\rm R} + G'_{\rm R}
+\tilde M_F(H F_{\rm L} + K G_{\rm L}) = 0
\ . \end{equation}
Here $\tilde \omega$ is the fermion eigenvalue $\omega$
in units of $M_W$
\begin{equation}
\tilde \omega = \frac{\omega}{M_W}
\nonumber \\
\ . \end{equation}

For $\xi=0$,
finite energy solutions of these equations
correspond to extrema of the Weinberg-Salam theory,
such as the sphaleron or the non-topological soliton.
The above set of equations can then also be obtained by
substituting the ansatz eqs.(12)-(17) into
the general equations of motion.

\subsection{\bf Boundary conditions}

The boundary conditions are chosen such that the energy density
and the energy both are finite.

At the origin the boson functions satisfy
the boundary conditions
\begin{equation}
f_A(0)-1=f_B(0)=H'(0)=K(0)=0
\ , \end{equation}
while the fermion functions satisfy
\begin{equation}
F_R(0)=F_L(0)=0
\ , \end{equation}
and
\begin{equation}
G_R(0)=c_R \ , \ \ G_L(0)=c_L
\ , \end{equation}
where $c_R$, $c_L$ are unknown constants,
subject to the normalization condition (26).

At infinity the gauge and Higgs field functions lie on a circle
\begin{equation}
f_A(\infty)+if_B(\infty)=\exp(i\theta(\infty)) \ , \ \
H(\infty)+iK(\infty)=\exp(i\frac{\theta(\infty)}{2})
\ , \end{equation}
and $\theta(\infty)$ is an unknown function of $\xi$.
Therefore we choose
the boundary conditions
\begin{equation}
f_A'(\infty)=f_B'(\infty)=H'(\infty)=K'(\infty)=0
\   \end{equation}
for the boson functions.
The fermion functions all vanish at infinity
\begin{equation}
F_R(\infty)=F_L(\infty)=G_R(\infty)=G_L(\infty)=0
\ . \end{equation}

\section{Minimal Energy Path}

We construct the minimal energy path
from one vacuum sector with $N_{\rm CS}=0$ to
the neighbouring vacuum sector with $N_{\rm CS}=1$
by solving the equations of motion (30)-(37),
varying the Chern-Simons number
along the path by means of the lagrange multiplier $\xi$.
For convenience we start at the top of the barrier,
at the sphaleron solution, where $\xi=0$,
and then vary the lagrange multiplier
$\xi$ to positive and to negative values
to follow the path down to the vacua on both sides
of the barrier.

\subsection{Sphaleron}

The sphaleron represents the configuration at the top of the barrier
between topologically distinct vacua.
It is a solution of the classical equations of motion
of the Weinberg-Salam theory with one unstable mode.
Previously the sphaleron solution was obtained only in
the bosonic sector of the Weinberg-Salam theory,
neglecting the fermions.
Here we investigate the influence which the presence of
fermions has on the sphaleron solution.

Without fermions, the sphaleron (for $\theta_{\rm w}=0$)
has spherical symmetry and parity reflection symmetry.
(We do not consider here the bisphalerons [23-24],
where parity reflection symmetry is broken,
and which arise at large values of the Higgs mass.)
Parity reflection symmetry simplifies the ansatz
for the bosonic fields of the sphaleron eqs.(12)-(14).
Two of the three gauge field functions, namely
$f_B$ and $f_C$, and one of the Higgs field functions, $H$, vanish
in the commonly used ansatz for the sphaleron.
The sphaleron then has a Chern-Simons number of precisely 1/2 [5].

Fermions have a zero mode
in the background field of the sphaleron.
For zero mass fermions this zero mode is known analytically
in terms of the sphaleron functions [11-13].
Recently this zero mode has
been constructed explicitly
for the case, where the fermions of a
weak doublet are degenerate in mass [16].
In this case two of the four fermion functions, namely
$\FR$ and $\FL$, vanish.

In a self-consistent treatment the presence of fermions affects
the bosonic fields through the coupled equations of motion.
The equation for the parity violating gauge field function $f_B$
now contains a source term from the fermions,
which does not vanish, even if $\FL$ were zero.
Similarly the equation for the previously vanishing
Higgs field function $H$ now contains a non-vanishing
source term from the fermions (except for zero mass fermions,
where $\GR=0$).
Thus the presence of fermions violates the parity reflection
symmetry of the sphaleron.
The previously vanishing parity violating functions
now have source terms from the fermions and can therefore
no longer vanish.

As a consequence the fermions couple to
a less symmetric bosonic configuration
and therefore also the previously vanishing fermion functions
$\FR$ and $\FL$ now have non-vanishing source terms,
even if the fermion eigenvalue would still be zero.
The self-consistent sphaleron solution thus
involves eight radial functions, four for the bosons
and four for the fermions.
With the parity reflection symmetry of the sphaleron lost,
the barrier no longer needs to be symmetric with respect
to the Chern-Simons number $N_{\rm CS}=1/2$.
Therefore the Chern-Simons number of
the sphaleron at the top of the barrier
may deviate from $1/2$,
and the eigenvalue of the fermions at the sphaleron
solution may deviate from zero.

In Fig.~1 we show the boson functions for the sphaleron
for the Higgs mass $M_H=M_W$
and for the fermion masses $M_F=130$ GeV and $M_F=1.3$ TeV for one
bound fermion.
We see, that even at a fermion mass of $M_F=130$ GeV,
corresponding to a mass on the order of the expected top quark mass,
the functions $f_B$ and $H$ are still almost vanishing,
and the deviation of the functions $f_A$ and $K$ from those
of the purely bosonic sphaleron is with only about 0.1\%
even an order of magnitude smaller.
In contrast, for heavy fermions with masses in the TeV region
$f_B$ and $H$ differ considerably from zero, while the deviation
of the functions $f_A$ and $K$
from those of the purely bosonic sphaleron is still small.

In Fig.~2 we show the corresponding fermion functions.
At the fermion mass of $M_F=130$ GeV
the functions $\FR$ and $\FL$ are still almost vanishing
in the self-consistent calculation,
they are smaller than $10^{-3}$,
while the functions $\GR$ and $\GL$ are almost identical
to those of the background field calculation.
For heavy fermions with masses in the TeV region
the functions $\FR$ and $\FL$ differ markedly from zero.

In Fig.~3 we show the energy of the sphaleron for three
values of the Higgs mass, $M_H=50,\ 80,\ 100$ GeV,
as a function of the fermion mass, for one and three bound
fermions, respectively.
For massless fermions
the energy of the self-consistent sphaleron
is almost unaltered
compared to the purely bosonic sphaleron.
For $M_H=M_W$ it increases by 0.0003 \% (0.0034 \%)
for one (three) bound fermion(s).
With increasing fermion mass the sphaleron energy
first increases slightly until about $M_F=120$ GeV
and then decreases for larger values of the fermion mass.
For a fermion mass on the order
of the top quark mass, $M_F= 130$ GeV,
the sphaleron energy decreases by only
0.003 \% (0.028) with respect to zero mass fermions
for $M_H=M_W$.
While for fermion masses approaching the TeV region,
the sphaleron energy decreases more strongly,
e.~g.~for a fermion mass of 1.3 TeV the energy changes
by 1.8 \% (12.5 \%) for one (three) bound fermion(s)
and $M_H=M_W$.

Fig.~4 shows the correponding energy in the boson fields
for the sphaleron as a function of the fermion mass.
Here we see an increase in energy for large fermion masses.
The decrease of the total energy for large values of the
fermions mass
must thus be due to the fermion eigenvalue,
which must decrease considerably
for large values of the fermion mass.
The fermion eigenvalue is shown in Fig.~5,
where we observe indeed a strong decrease for heavy fermions.
Thus the fermion eigenvalue of the self-consistent sphaleron
deviates strongly from zero for heavy fermions.

For light and intermediate fermions
the eigenvalue is shown in Fig.~6.
Here we observe small positive values of the eigenvalue.
At a fermion mass of zero the eigenvalue starts
at 0.07 GeV (0.2 GeV) for one (three) bound fermion(s),
rises to a maximum of
0.6 GeV (1.8 GeV)
for a fermion mass on the order of the top quark mass,
and then declines, passing zero
for a fermion mass of about 0.2 TeV
in both cases for $M_H=M_W$.

In Fig.~7 the Chern-Simons number of the sphaleron solution
is shown for the same set of Higgs masses and for one (three)
bound fermion(s) as a function of the fermion mass.
For zero mass fermions the Chern-Simons number
approaches 0.499 (0.497) for one (three) bound fermion(s).
For fermions with mass $M_F=130$ GeV the corresponding
Chern-Simons number is 0.495 (0.485),
while for heavy fermions with mass $M_F=1.3$ TeV
the Chern-Simons number changes by 10 \% (31 \%)
for $M_H=M_W$.
Thus the Chern-Simons number differs appreciably
from the value $N_{\rm CS}=1/2$
only for heavy fermions.

\subsection{Barrier}

When the barrier is calculated self-consistently
in the presence of fermions
the barrier is no longer
symmetric with respect to $N_{\rm CS}=1/2$.
The minimal energy path
from the sphaleron down to the vacuum sector with
$N_{\rm CS}=0$ is shorter and steeper
than the path down to the other side to the
vacuum sector with $N_{\rm CS}=1$.

The minimal energy path is shown in Fig.~8
for $M_H=M_W$ for one light fermion with $M_F=1/10 M_W$
and for one (three) heavy fermion(s) with $M_F=10 M_W$.
For the light fermion the path
agrees well with the purely bosonic path [14].
The deviation is very small at the top
and increases slightly along the path
towards both vacuum sectors.
There the positive mass of the fermions
is approached for $N_{\rm CS} \rightarrow 0$
and the negative mass for $N_{\rm CS} \rightarrow 1$,
because there is now a free fermion present
with positive or negative energy.

For large values of the fermion mass,
the barrier is altered considerably.
With increasing fermion mass
the top of the barrier moves to smaller values
of the Chern-Simons number,
making the barrier increasingly asymmetric.
Beyond a fermion mass of about 0.7 GeV
for one bound fermion
the minimal energy path heading towards the
$N_{\rm CS} = 0$ sector
does not reach the bosonic vacuum configuration
with one (three) free fermion(s) any more.
Instead a new bound state is reached,
a non-topological soliton [19-21], discussed below.
The minimal energy path down to the other side of the sphaleron
towards the $N_{\rm CS} = 1$ sector
does reach a bosonic vacuum configuration
with one (three) free fermion(s) with negative energy.

In Fig.~9 we show the dependence of the Chern-Simons number
on the lagrange multiplier $\xi$ along the minimal energy path
for $M_H=M_W$, for one light fermion with $M_F=1/10 M_W$
and for one (three) heavy fermion(s) with $M_F=10 M_W$.
For small fermion masses the relation is
monotonic and almost linear.
While $N_{\rm CS}$ varies from zero to one,
the lagrange multiplier $\xi$ changes from -2 to +2.
In contrast for heavy fermions,
where non-topological soliton solutions exist,
$\xi$ reaches a minimum close to -2,
and then rapidly increases again,
approaching zero at the non-topological soliton solution
as shown in Fig.~10. (Note, that the Chern-Simons number
of the non-topological soliton is nonzero.)

In Fig.~11 we show the fermion eigenvalue $\omega$ along the
minimal energy path in the self-consistent calculation
for $M_H=M_W$ for one light fermion with $M_F=1/10 M_W$
for one intermediate fermion with $M_F= M_W$,
and for one (three) heavy fermion(s) with $M_F=10 M_W$.
Comparing the eigenvalues obtained in the self-consistent
calculation with those of
the background field calculation [16]
we see little change for light and intermediate fermions.
The eigenvalue still looks (almost) antisymmetric
with respect to $N_{\rm CS}=1/2$.
In contrast, for heavy fermions
the self-consistent minimal energy path starts from
the non-topological soliton
and thus with an eigenvalue $\omega < M_F$.
The eigenvalue decreases monotonically
with increasing Chern-Simons number and reaches
zero long before the sphaleron is reached
(e.~g.~for $M_F=0.8$ TeV
the fermion mode has eigenvalue zero at $N_{\rm CS}=0.4354$,
while the sphaleron is reached at $N_{\rm CS}=0.4700$
for $M_H=M_W$).
Then the eigenvalue gradually approaches the negative
mass, $\omega \rightarrow - M_F$.

There exist critical values of the fermion mass where,
at a given value of the Chern-Simons number $N_{\rm CS}$,
the bound state enters the positive or the negative continuum.
For light and intermediate mass fermions
the critical values of the self-consistent
calculation almost coincide with those
of the background field calculation [16].
For instance for $M_F = 80$ GeV
the bound state enters the positive continuum
at $N_{\rm CS}=0.0773$ and the negative continuum at
$N_{\rm CS}=0.9216$ in the self-consistent calculation
for one fermion,
and at $N_{\rm CS}=0.0748$
and $N_{\rm CS}=0.9252$ in the background field calculation [16].
For heavy fermions
the self-consistent solution reaches the non-topological soliton
in the $N_{\rm CS}=0$ sector,
leading to a very different behaviour for
$N_{\rm CS} \rightarrow 0$.

In Fig.~12 we show the value of the Chern-Simons number
where, along the minimal energy path,
the fermions have eigenvalue zero
for $M_H=M_W$ and one (three) bound fermion(s).
For massless fermions the eigenvalue is zero
precisely at a Chern-Simons number of 1/2.
(The contributions to $N_{\rm CS}$ from the integral
and from the angle $\theta(\infty)$
in eq.(29) cancel to better than $10^{-5}$.)
The Chern-Simons number of the zero mode
then decreases with increasing fermion mass
rapidly towards zero.
The end configurations
correspond to non-topological solitons
with zero energy eigenstates, which are reached
at $M_F=2.416$ TeV ($1.384$ TeV)
with $N_{\rm CS}=0.00038$ ($0.00038$)
for one (three) bound fermion(s).

Let us now compare the self-consistent path
with the path obtained in [19].
For very heavy fermions the self-consistent minimal energy path
no longer decreases monotonically from the sphaleron towards the vacua.
Instead a new critical behaviour appears
for fermion masses around 4 TeV.
The path down to the vacuum sector $N_{\rm CS}=0$
turns backward at a first critical point
and forward again at a second critical point.
This is illustrated in Fig.~13 for the fermion masses
$M_F=4$ TeV, $M_F=6$ TeV, and $M_F=10$ TeV for $M_H=M_W$.
In Fig.~14 the correponding dependence of the lagrange
multiplier $\xi$ on the Chern-Simons number $N_{\rm CS}$
is shown. While for $M_F=4$ TeV and $M_F=6$ TeV
there is only one configuration with $\xi=0$,
there are three such configurations for
$M_F=10$ TeV. Since all configurations with $\xi=0$
correspond to solutions of the classical field
equations of the (simplified) Weinberg-Salam theory,
there are additional unstable solutions for these
high fermions masses, new sphalerons.

In [19]
a restricted variational ansatz is used for the
boson functions, which approach the vacuum configuration
on both sides of the barrier,
while the fermion functions are
obtained perturbatively (in $f^{(q)}/g$),
and the fermion energy approaches the positive and
negative free mass.
Thus this variational calculation
does not approach the non-topological soliton.
Not surprisingly this variational ansatz leads to a very
different dependence of the barrier on the fermion mass.
When the fermion mass runs across the TeV region
critical values of the fermion mass are encountered,
where a new local energy minimum first appears and
then disappears again along path [19].
The new local minimum is associated with a new kind of soliton in [19],
different from the non-topological soliton [19-21],
and existing only for very heavy fermions
with 9.0 TeV $\le M_F \le 12.4$ TeV for $M_H=M_W$.

There are several reasons to doubt the argument
for the existence of a new soliton, given in [19].
The calulation of [19] is only variational
and therefore gives only an upper bound on the energy.
Our calculation shows, that no such new smooth
minimum appears along the self-consistent path,
although a different critical behaviour does appear
for fermion masses around 4 TeV.
Furthermore, for fermion masses in the critical range
9.0 TeV $\le M_F \le 12.4$ TeV
the non-topological soliton has a negative energy,
indicating the breakdown of the valence
fermion approximation [20,21].

\subsection{Non-topological soliton}

We have seen, that
for heavy fermions one encounters a non-topological
soliton along the minimal energy path when approaching
the vacuum sector with $N_{\rm CS}=0$.
We now take a closer look at this solution of the
Weinberg-Salam model, which represents a minimum
and not a saddle point like the sphaleron.

Since the gauge field has little influence on the
non-topological soliton [19],
we first consider a simpler model by setting the gauge field
equal to zero. In this case the equations for
the left-handed and for the right-handed fermions
are equivalent and we are essentially left with a
chiral linear $\sigma$-model, scaled up in energy
since the vacuum expectation value of the Higgs field
is about 25 times bigger than $f_{\pi}$.

In Fig.~15 we show the total energy
of the $\sigma$-model non-topological soliton
as a function of the fermion mass for
one bound fermion and
for several values of the Higgs mass
ranging from 50 GeV to 5 TeV.
We see, that for a given Higgs mass,
there exists a critical value of
the fermion mass above which
the fermion can form a stable bound state, the
non-topological soliton, which is
lower in energy than the mass of one free fermion.
These results agree with those of [19],
where a direct minimization technique was used.

For a given fermion mass
the energy of the non-topological soliton
increases monotonically with the Higgs mass.
Thus the energy is clearly bounded from above by the
energy obtained in the non-linear $\sigma$-model,
where the Higgs field is confined to the chiral circle [19-21].
The energy of the solitons becomes negative
for fermion masses in the range $3.6 \le M_F \le 3.8$ TeV
for the Higgs masses considered.
This behaviour is largely due to the valence fermion
approximation, since taking into account the effects of
the Dirac sea leads to a qualitatively different behaviour [20-21].

We observe that the soliton branches
approach the free fermion branch
in two distinct ways, depending on
the value of the Higgs mass.
For low Higgs masses the energy along the soliton branch
is always lower than the corresponding free energy,
and the soliton branch simply bifurcates from
the free fermion branch at a critical value of the fermion mass.
Above a critical Higgs mass of about $M_H =0.7$ TeV
for one bound fermion, however,
the behaviour is different. Here the soliton branch
first crosses the free branch,
then reaches a cusp at a critical value of the fermion mass,
where a new (unstable) branch emerges.
The new branch then approaches the free branch
from above. This latter behaviour is well known for
non-topological solitons [25].

The different critical behaviour of the soliton branches
at various Higgs masses is also seen in Fig.~16.
There we show the fermion eigenvalue as a function
of the fermion mass for the same set of
values of the Higgs mass.
When for large values of the Higgs mass the energy branch approaches
the cusp, the fermion eigenvalue exhibits an infinite slope.

In Fig.~17 we show the radial functions
for the non-topological soliton for $M_H=M_W$ and
$M_F=2.6$ TeV for one bound fermion.
We see, that the fermions are localized in a small region of
space, while the Higgs field approaches its vacuum
expectation value, where $H(x)\rightarrow 1$
and $K(x)\rightarrow 0$, much more slowly.

Finally we consider the effect of the presence of the gauge
field on the non-topological solitons
in the Weinberg-Salam model.
In Fig.~18 we show the gauge field functions
of the non-topological soliton
for the physical gauge coupling $g=0.65$
for $M_H=M_W$ and $M_F=1.3$ TeV and
$M_F=2.6$ TeV for one bound fermion.
They differ only little from zero.

The change in energy of the non-topological soliton
due to the presence of the gauge field
is smaller than 0.8 \% for $M_F=2.6$ TeV and
for $M_H=M_W$.
The deviation decreases for smaller fermion masses.
The non-topological solitons
possess a small, but finite Chern-Simons number,
$N_{\rm CS} \le 0.0004$
for $M_F \le 2.6$ TeV and $M_H=M_W$.

\section{Conclusions}

In the Weinberg-Salam theory topologically distinct vacuum sectors
are seperated by a barrier, whose height is determined
by the sphaleron energy.
Here we have constructed the minimal energy path
over the barrier from one vacuum sector
to the neighbouring one self-consistently
in the valence fermion approximation.
To retain spherical symmetry we have neglected the weak mixing angle
and assumed that the fermions of a doublet are degenerate
in mass. While the first approximation is very good [17,18],
the latter is badly broken and needs to be improved.

The presence of the fermions affects the energy and the shape of the
barrier. The sphaleron solution on top of the barrier
loses its parity reflection symmetry due to the coupling
to the fermions. It carries no longer exactly Chern-Simons number
$N_{\rm CS}=1/2$, and the fermion eigenvalue at the top of the
barrier is no longer precisely zero.
The deviations of the self-consistent sphaleron from the
purely bosonic sphaleron depend on the mass
and on the number of bound fermions.
The presence of massless fermions hardly affects the
sphaleron, while fermions with masses on the order
of the top quark mass have a small but noticeable effect.
The Chern-Simons number is decreased by about one percent
from $N_{\rm CS}=1/2$,
and the fermion eigenvalue differs slightly from zero,
being on the order of one percent of the fermion mass.
Though the height of the barrier is still insensitive
in this mass range.
Due to these small changes the barrier is very slightly
tilted to the left, becoming asymmetric with respect
to the top. The fermion eigenvalue along the self-consistent path
differs little from
the one of the background field calculation [16]
for light and intermediate mass (top quark) fermions.
We conclude that in this mass range the sphaleron and the
barrier are well approximated by neglecting the fermions,
and that the fermion eigenvalues are obtained with rather
good accuracy in the background field calculation,
as compared to the self-consistent valence fermion
approximation.
Clearly other approximations, such as neglecting
the effect of the Dirac sea, still need to be investigated.

For heavy fermions with masses approaching the TeV region
the barrier deforms stronger.
The height of the barrier then decreases monotonically
with increasing fermion mass,
and decreases the stronger the more fermions occupy the
valence level.
Likewise the Chern-Simons number of the sphaleron
and the fermion eigenvalue at the sphaleron decrease.

Heavy fermions with masses greater than about 0.7 TeV
can form non-topological solitons in the vacuum sector.
For such heavy fermions the self-consistent minimal energy path
over the barrier does not end in the vacuum configuration
but instead reaches the non-topological soliton
as its end configuration. This is in contrast to
the variational approximate path considered
by Petriashvili [19], which for any value of the
fermion mass ends in a vacuum with a free fermion.
Due to this choice of end configuration Petriashvili
finds a new minimum along the path for very heavy fermions,
and therefore conjectures the existence of
a new type of soliton [19].
Our self-consistent calculations also show a new critical
behaviour for very heavy fermions. But we do not observe
a new smooth minimum along the path. Instead we see the occurrence
of two bifurcations for very heavy fermions.
The minimal energy path then encounters two critical points.
It winds backward at the first critical point,
and forward again at the second critical point.
This critical behaviour allows for the existence of
additional unstable solutions, new sphalerons.
Our self-consistent calculation does
not support the conjecture, that a new type of soliton
is present for very heavy fermions.

{\bf Acknowledegement}

We gratefully acknowledge discussions with Yves Brihaye and
Burkhard Kleihaus.

\newpage

\section{Figure Captions}

\indent{\bf Figure 1:}
The gauge field functions $f_A$ and $f_B$
and Higgs field functions $H$ and $K$
of the sphaleron are shown
with respect to the dimensionless variable x
for $M_H=M_W$
and for $M_F=130$ GeV (solid) and $M_F=1.3$ TeV (dotted).

{\bf Figure 2:}
The righthanded and lefthanded fermion functions
$G$ and $F$
of the sphaleron are shown
with respect to the dimensionless variable x
for $M_H=M_W$
and for $M_F=130$ GeV (solid and dot-dashed)
and $M_F=1.3$ TeV (dotted and dashed).

{\bf Figure 3:}
The total energy (in TeV)
of the sphaleron is shown as a function of
the fermion mass (in TeV) for three values of the Higgs mass,
$M_H=50$ GeV, $M_H=80$ GeV, and $M_H=100$ GeV,
for one fermion (solid) and three fermions (dotted).

{\bf Figure 4:}
The boson energy (in TeV)
of the sphaleron is shown as a function of
the fermion mass (in TeV) for three values of the Higgs mass,
$M_H=50$ GeV, $M_H=80$ GeV, and $M_H=100$ GeV,
for one fermion (solid) and three fermions (dotted).

{\bf Figure 5:}
The fermion eigenvalue (in TeV)
of the sphaleron is shown as a function of
the fermion mass (in TeV) for three values of the Higgs mass,
$M_H=50$ GeV (solid), $M_H=80$ GeV (dashed),
and $M_H=100$ GeV (dot-dashed),
for one fermion and three fermions.

{\bf Figure 6:}
The fermion eigenvalue (in GeV)
of the sphaleron is shown as a function of
the fermion mass (in TeV)
for light and intermediate mass fermions
for three values of the Higgs mass,
$M_H=50$ GeV (solid), $M_H=80$ GeV (dashed),
and $M_H=100$ GeV (dot-dashed),
for one fermion and three fermions.

{\bf Figure 7:}
The Chern-Simons number
of the sphaleron is shown as a function of
the fermion mass (in TeV) for three values of the Higgs mass,
$M_H=50$ GeV (solid), $M_H=80$ GeV (dashed),
and $M_H=100$ GeV (dot-dashed),
for one fermion and three fermions.

{\bf Figure 8:}
The energy (in TeV) along the minimal energy path
is shown as a function of
the Chern-Simons number for $M_H=M_W$
for one fermion with $M_F=1/10 M_W$ and $M_F=10 M_W$ (solid)
and for three fermions with $M_F=10 M_W$ (dotted).

{\bf Figure 9:}
The lagrange parameter $\xi$ along the minimal energy path
is shown as a function of
the Chern-Simons number for $M_H=M_W$
for one fermion with $M_F=1/10 M_W$ and $M_F=10 M_W$ (solid)
and for three fermions with $M_F=10 M_W$ (dotted).

{\bf Figure 10:}
The lagrange parameter $\xi$ along the minimal energy path
is shown in the vicinity of the non-topological soliton
as a function of
the Chern-Simons number for $M_H=M_W$
for $M_F=10 M_W$
for one (solid) and three (dotted) fermions.

{\bf Figure 11:}
The eigenvalue (in units of $M_F$) along the minimal energy path
is shown as a function of
the Chern-Simons number for $M_H=M_W$
for one fermion with $M_F=1/10 M_W$, $M_F=M_W$
and $M_F=10 M_W$ (solid)
and for three fermions with $M_F=10 M_W$ (dotted).

{\bf Figure 12:}
The Chern-Simons number of the zero mode
along the minimal energy path
is shown as a function of
the fermion mass (in TeV)
for $M_H=M_W$
for one fermion and three fermions.

{\bf Figure 13:}
The energy (in TeV)
along the minimal energy path
is shown as a function of
the Chern-Simons number for $M_H=M_W$
for one heavy fermion with $M_F=4$ TeV,
$M_F=6$ TeV, and $M_F=10$ TeV.
(The dotted parts are extrapolated.)

{\bf Figure 14:}
The lagrange multiplier $\xi$
along the minimal energy path
is shown as a function of
the Chern-Simons number for $M_H=M_W$
for one heavy fermion with $M_F=4$ TeV
$M_F=6$ TeV, and $M_F=10$ TeV.
(The dotted parts contain non-negligible numerical errors.)
The curves cross zero at sphaleron solutions.

{\bf Figure 15:}
The energy (in TeV) of the non-topological soliton
for vanishing gauge coupling constant
is shown as a function of the fermion mass (in TeV)
for $M_H=80$ GeV (dot-dashed), $M_H=0.5$ TeV (dashed),
and $M_H=1$ TeV (dotted) and $M_H=5$ TeV (solid)
for one bound fermion.

{\bf Figure 16:}
The fermion eigenvalue (in TeV) of the non-topological soliton
is shown as a function of the fermion mass (in TeV)
for $M_H=80$ GeV (dot-dashed), $M_H=0.5$ TeV (dashed),
and $M_H=1$ TeV (dotted) and $M_H=5$ TeV (solid)
for one bound fermion
for vanishing gauge coupling constant.

{\bf Figure 17:}
The fermion field functions
$G$ (solid) and $F$ (dotted)
and Higgs field functions $H$ (dashed) and $K$ (dot-dashed)
of the non-topological soliton
are shown with respect to the dimensionless variable x
for $M_H=M_W$ and for $M_F=2.6$ TeV
for vanishing gauge coupling constant.

{\bf Figure 18:}
The gauge field functions
$1-f_A$ and $f_B$ of the non-topological soliton are shown
with respect to the dimensionless variable x for $M_H=M_W$
and for $M_F=2.6$ TeV (solid) and $M_F=1.3$ TeV (dotted)
for the physical value of the gauge coupling.

\end{document}